# Can AI Moderate Online Communities?

Henrik Axelsen, Johannes Rude Jensen, Sebastian Axelsen, Valdemar Licht, Omri Ross

*Work in Progress*

## Abstract

The task of cultivating healthy communication in online communities becomes increasingly urgent, as gaming and social media experiences become progressively more immersive and life-like. We approach the challenge of moderating online communities by training student models using a large language model (LLM). We use zero-shot learning models to distill and expand datasets followed by a few-shot learning and a fine-tuning approach, leveraging open-access generative pre-trained transformer models (GPT) from OpenAI. Our preliminary findings suggest, that when properly trained, LLMs can excel in identifying actor intentions, moderating toxic comments, and rewarding positive contributions. The student models perform above-expectation in non-contextual assignments such as identifying classically toxic behavior and perform sufficiently on contextual assignments such as identifying positive contributions to online discourse. Further, using open-access models like OpenAI's GPT we experience a step-change in the development process for what has historically been a complex modeling task. We contribute to the information system (IS) discourse with a rapid development framework on the application of generative AI in content online moderation and management of culture in decentralized, pseudonymous communities by providing a sample model suite of industrial-ready generative AI models based on open-access LLMs.

## Introduction

Moderating online community interactions presents an important challenge, given the increasingly immersive nature of the in-game experience. As the quality of metaverse gaming experiences increases, the task of moderating user interactions becomes increasingly important. As gamers submerge themselves in increasingly sophisticated metaverses, they become vulnerable to traumatizing interactions.

The IS literature have shown that moderation directly affect participants incentive to follow guidelines and produce meaningful contributions (Chen et al. 2011). It is argued that purposefully moderated communities can stimulate user participation and advance constructive discourse, in the presence of machine-*assisted* moderation (He et al. 2020). Yet, while recent, these findings do not factor in the recent emergence of highly performant large language models (LLM) such as the highly publicized ChatGPT.

ChatGPT is a large language model based on the GPT-3.5 architecture, developed, and trained by OpenAI, which has gained significant attention in the field of Natural Language Processing (NLP) due to its ability to generate human-like responses and carry out a variety of tasks such as text completion, translation, and summarization.

Given recent advancements in conversational abilities exhibited by modern access LLMs, we hypothesize that these models may outclass previous attempts at moderating online communities. If successful, the task of cultivating a safe environment for healthy and productive discourse online, is within reach. We address the research question: "To what extent can LLMs be used in moderating online communities?".

We present preliminary results from an ongoing case study, in which we attempt to automate the task of content moderation in a small online community for a metaverse game, by using the suite of Generative Pre-trained Transformer (GPT) LLMs made available by OpenAI. Through a distillation and curation process, we train three 'student-models' aimed at (I) identifying the intent of pseudonymous actors (II) identifying and moderating toxic behavior and (III) curating and promoting novel contributions, in online communities. Using a process referred to as 'knowledge distillation', student models distill general knowledge from the 'teacher' to the smaller student model, which is trained on a significantly smaller corpus. The student model can subsequently be tuned to optimize its performance for a specific context, improving the model's ability to perform a set of objectives within a specific context.

Our preliminary results indicate that when properly trained, existing open-access GPT LLMs, such as GPT3, excel in understanding discursive trends in online subcultures. We demonstrate how the three student models produce exceptionally well on classical tasks such as intent and moderation, while falling short of expectations on contextual assignments such as classifying positive contributions to the community. These findings suggest that, given proper training of the models, mass content moderation can be made more effective through advanced analytics, and that analytics can successfully monitor and moderate discourse in online fora.

While the IS literature on data science and AI displays a broad range of perspectives on the interface between human and machine, we note the relative scarcity of conceptual research addressing practical challenges. This short paper contributes practical results towards furthering our understanding of the limitations and challenges in applying artificial intelligence in practical use cases. Moreover, we contribute to the literature on the ability of analytics and algorithms to resolve societal issues such as online mediated polarization through proper moderation.

## Literature Review

While the release of the OpenAIs GPT model has taken the world by storm, the groundbreaking implications of AI have been predicated in the IS literature years prior. The IS literature spans a wide variety of topics concerning the implications of AI. Particularly, IS scholarship examines the interface between humans and machines, notably in sensitive situations in which an AI agent is tasked with replacing humans as the counterparty in business or personal conversations (Benbya et al. 2020). Empirical data suggest that the level of trust imbued by humans in AI agents is a factor of the perceived accuracy of the results (Salem et al. 2015) and, equally important, the perceived anthropomorphism of the model and it's reasoning process (Glikson and Woolley 2020).

Schanke and colleagues (Schanke et al. 2021) examines this issue in the degree at which "humanizing" customer service chatbots affects customer satisfaction with their experience and their opinion the company. The authors demonstrate that higher degrees of perceived anthropomorphism correlates with higher levels of satisfaction.

Conversational agents such as chatbots are considered more anthropomorphic by humans when they appear "human-like" in their speech. For example, they might express what is perceived as emotions or humor to compel the subject (Seeger et al. 2021).These findings have led authors to argue that customer or user-facing bots should be used in tandem with human support, augmenting the capacity of humans, but not replacing them entirely (He et al. 2020). Designers should consider carefully the extent to which bots should appear human-like, as humans may quickly come to trust the bot with responsibilities that are not appropriate for algorithms (Seeger et al. 2021). Because algorithms may never become able to fully grasp the complexity of emotional cues in conversation, their capacity to respond appropriately in out-of-context emergency situations is severely limited (Purdy et al. 2019).

These warnings stand out as pertinent in the light of the extant literature in the field. Two critical issues in the implementation of bot-assisted human judgement in conversational tasks such as forum moderation is complacency and overreliance on the model, which can serve to reinforce bias. Humans may become complacent and overly reliant on algorithmic support, which can cause them to fail in noticing errors or incorrect information (Parasuraman and Manzey 2010). This brings to the forefront the emerging power dynamics between human and machine, as algorithmic decision-making may gradually undermine the values that underpin organizations (Lindebaum et al. 2020) as it optimizes for other values.

These concerns are perhaps no more evident than in the perpetuation of implicit bias. As AI technology are, by definition, trained on retrospective data they are likely to perpetuate a myopic and 'conservative' view on the world, unless trained otherwise (Balasubramanian et al. 2022). While LLMs can be trained to understand a range of progressive values, empirical indicators for implicit associations and beliefs perpetuated through unconscious linkages between phenomena such as "technology" and "masculinity" are much harder to address (Elsbach and Stigliani 2019).

These warnings have led to suggestions for carefully monitored implementation of AI in organizations, with arguments that AI should be considered 'augmenting automation' that complements rather than replaces human workers (Tschang and Almirall 2021). Especially when systems are hard to understand, they should be enveloped within a set of clear set of operational boundaries (Asatiani et al. 2021) and contextually applied to support and alleviate human activities.

## Research Design

### *Case Descriptions and Data Collection*

We draw our training data from Reality+, a metaverse gaming designer operating a gaming platform utilizing non-fungible tokens (NFTs) as part of the in-game experience (Regner and Schweizer 2019). Reality+ offers multiple gaming experiences, all of which utilize NFTs to represent ownership of items of property inside the game. In parallel, gamers utilize the popular server *Discord* to communicate. We draw data specifically from the online metaverse game "Doctor Who Worlds Apart" (DWWA), in which gamers emulate scenes from the popular British TV series, collecting NFTs in the process. Of ~250,000 blockchain wallets connected to Reality+'s NFT platform, ~75,000 are related to DWWA. Nevertheless, only a small fraction of the connected wallets currently constitutes the gaming community. Of the approximately ~3,000 active users in the game, only ~1,100 participants are active in the chat at the time of writing. The DWWA community is managed by some 30 moderators.

Yet, in recent months, growth has accelerated and with a planned hypergrowth strategy rollout of a device-agnostic multimedia gaming experience based on a novel contribution model there is a high risk that moderators become overburdened with the task of managing online discourse in the metaverse experience, as the data-set revealed an unimpressive Krippendorff alpha coefficient of 0.254 among a panel of 4 humans (Krippendorff 2018) indicating poor consistency in human interpretation of cultural practices of the community.

*Method*

We explore an LLM based model approach to content moderation as an alternative to resource-intensive human moderation, which has historically posed significant challenges at scale. There are several ways to model the problems analytically. All boil down to a question of data availability, performance, and cost (Table 1.).

| Method | Advantages | Drawbacks |
| --- | --- | --- |
| Train our own natural language classifier using word embeddings and Long Short Term Memory (LSTM) architecture. | Complete control over the modeling process, low-cost and low-latency inference. | Requires a large training corpus, requires training hardware (e.g., GPU machines), and the approach highly depends on feature engineering (how good the embeddings are). |
| Use a zero-shot LLM such as OpenAI's text-davinci model. | Entirely API-based (no specialized hardware requirements), very easy to get started with, requires no training corpus. | Likely requires a powerful model for sufficient results, expensive, high latency inference, highly dependent on prompt engineering. |
| Few-shot learning or fine-tune an existing LLM such as OpenAI's ada or davinci models. | Entirely API-based, easy to start with, and requires only a small training corpus. | Requires building a training corpus. Higher latency and inference cost than designing from scratch. |
| **Table 1. Modeling approaches** | | |

We chose OpenAI's GPT3 model suite for the task, as it offers a range of large general-purpose language models with different levels of sophistication and complexity. This makes it suitable for many text-related tasks like extraction, labeling, and classification. Additionally, OpenAI's GPT-3 models can be accessed through an API, providing a cost-effective and speedy way to prototype models. Most of the cost is incurred when the models are put into production, delaying the cost-benefit analysis until a later stage.

Where it was previously necessary to fine-tune LLMs on datasets of thousands of examples for good performance, with GPT3 scaling a language model requires only a few examples (few-shot learning) to achieve similar performance. (Brown et al. 2020). Alongside few-shot modeling, the deep unsupervised learning research field has made significant advancements with generative and embedding models using Contrastive Pre-Training (Neelakantan et al. 2022; Zhou et al. 2022) to generate text.

Our approach to model development involves using zero-shot learning models to distill data, when we lack data to validate with humans, followed by a few-shot learning or fine-tuning approach to reduce development time and cost. We generate data using zero-shot learning and fine-tune on this after having corrected the labels with a human annotator.

We start with an initial dataset and use iterative loops of fine-tuning to generate more annotated examples. These examples are continuously validated through moderators' inductive reasoning, mirroring traditional ways of taxonomy development and classification. The end result is few-shot models that are of lower complexity and cost (Nickerson et al. 2008). We found that the GPT-3 LLMs ada, babbage, curie, davinci, and gpt-3.5-turbo models are all sufficiently strong base models for this approach.

To protect intellectual property, Generative Pre-trained Transformer (GPT) models, of which OpenAI's GPT 3.5 and GPT4 are the leading instruction-following models available, are closed-source. The increasing complexity of these models and lack of available open-source access has led to a proliferation of model compression and acceleration techniques. The process of 'knowledge distillation' is a representative type of such a model compression and acceleration technique, in which a larger model effectively learns a small student model (Gou et al. 2021). This makes it possible to create more cost-efficient models for tasks requiring less complexity in production while achieving the benefits of much larger models ("Stanford CRFM" 2023). Our process follows such a distillation process, we designed the following procedural steps to rapidly create a suite of models based on the open-access API:

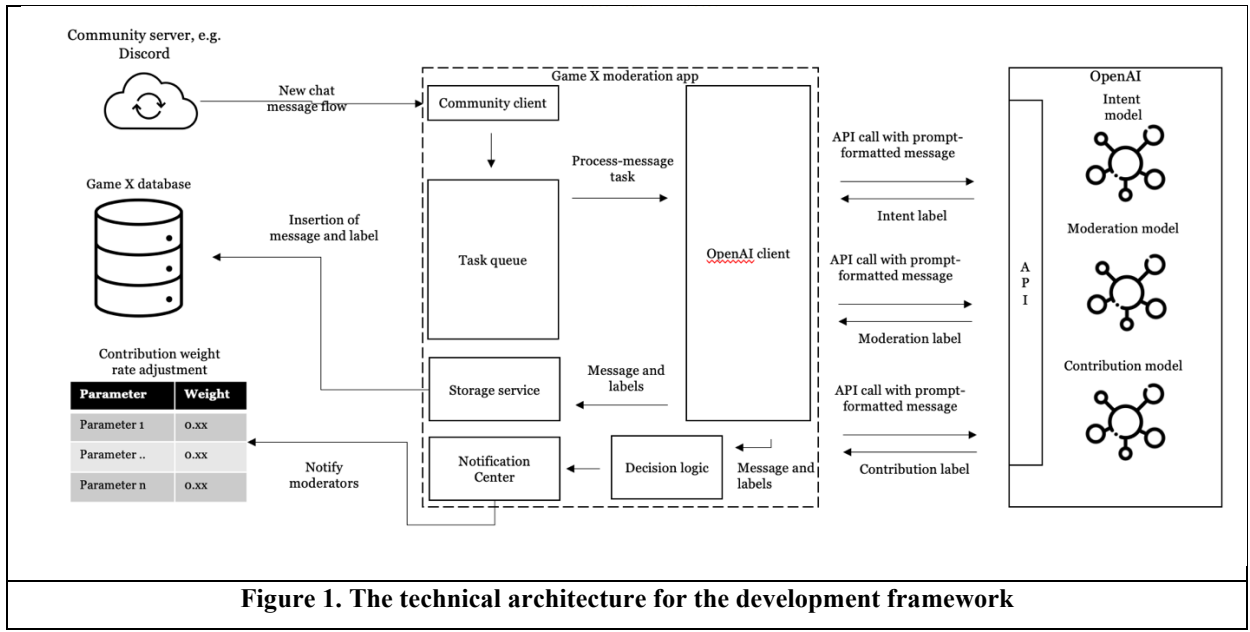

**Figure 1. The technical architecture for the development framework**

The architecture is used in an iterative 8-step process, in which prompting is used to classify messages and subsequently benchmark against the initially labeled data (figure 2). The model is used recursively (repeating step 4-6) until performance gains level out for the simplest, most cost-efficient student model.

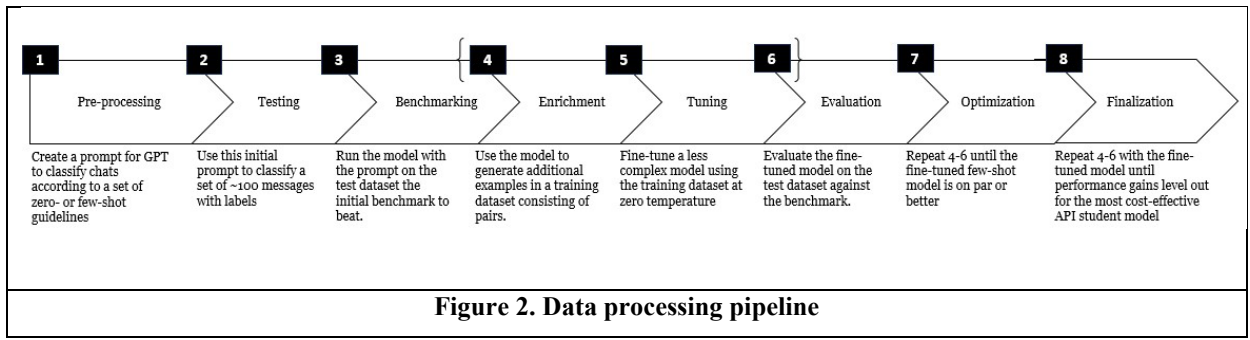

**Figure 2. Data processing pipeline**

If the model fails to perform past step 7, it may be that the problem is not a classification problem. If this is the case, first, we (the authors) let more than one moderator classify and validate the dataset to check the consistency of human annotation. This step introduces concern since human graders bring their biases to the task in addition to the potential bais inherent in the LLMs. As the task is classification, we use multiple graders and consensus in the curation process

## Preliminary Results

Using the architecture and processing pipeline, we train student-models for actor intent, moderation, and contribution. We score the models on *precision,* the model's ability to identify the right data-points. *Recall,* the model's ability to identify true positives and calculate the *F1 score,* as the weighted average between precision and recall (the harmonic mean).

*Actor Intent model*

Using the Doctor Who community as a baseline, we were supplied with Discord chat dataset consisting of over 60,000 chat messages. From these, we fine-tuned an OpenAI ada model, the smallest model, to generate an intent label given a chat message based on 463 human-labeled chat messages. The model performance was benchmarked against an out-of-sample test dataset of 116 human-labeled chat messages.

| Classifier | Precision | Recall | F1-score | *n* messages |
|---|---|---|---|---|
| Crypto | 0.92 | 0.80 | 0.86 | 41 |
| Fan | 0.93 | 1.00 | 0.96 | 40 |
| Casual | 0.86 | 0.91 | 0.89 | 35 |
| Accuracy | | | 0.91 | 116 |
| Macro avg across | 0.90 | 0.91 | 0.90 | 116 |
| Weighted avg | 0.91 | 0.91 | 0.90 | 116 |

**Table 2. Intent model results**

The model performance results (Table 2.) prove production ready after three iterations of fine-tuning with an estimated model development time of 18 hours over two weeks to allow for processing time for the manual labeling of the 463 chat messages. After fine-tuning the intent model, we ran it on the full database of 66,000 messages scraped from the Discord community, initially supplied by the moderator team. Excluding all empty messages and those from well-known bot accounts, this accounted for a total of 59,910 messages across 1,121 active users in 10 channels. The full loop took 6 hours to complete, and results showed that approx. 52% of the entire Discord conversation was casual chatter, 25% was related to the gaming universe, and approx. 18% was related to crypto. Categorizing users with three or more messages classified as "crypto" as Crypto Enthusiast personas, users with three or more messages classified as "fan" as Fan personas, and users who are neither Crypto Enthusiasts nor Fans as Casual personas, suggest that 343 unique (pseudonymous) IDs, or app. 31% of the active community were Crypto Enthusiasts, 243 or 22% Fans and 716 or 64% Casuals.

*Actor Moderation model*

OpenAI's GPT suite already includes a moderation model allowing moderation of toxic content related to hate, violence, sex, and threats, which is free to use when monitoring the inputs and outputs of OpenAI APIs, providing an additional benefit to the user. However, when testing the default model we experienced significant Type I errors (false-positives) related to gaming, such as 'this character will exterminate that character', or blockchain language, such as 'burning tokens', to be classified as violence. Further, the default moderation model in GPT does not flag spam, a common feature in Discord fora. So, we trained a GPT model using the same approach as above, leveraging past moderated messages in combination with a public dataset available from Kaggle's "Toxic Comment Classification Challenge."

| Classifier | Precision | Recall | F1-score | *n* messages |
|---|---|---|---|---|
| Toxic | 0.95 | 0.99 | 0.97 | 106 |
| Spam | 1.00 | 0.89 | 0.94 | 9 |
| Not_toxic_not_spam | 0.99 | 0.98 | 0.99 | 268 |
| Accuracy | | | 0.98 | 383 |
| Macro avg | 0.98 | 0.95 | 0.97 | 383 |
| Weighted avg | 0.98 | 0.98 | 0.98 | 383 |
| Table 3. Moderation model results | | | | |

Given the use case of this moderation tool, we want to minimize Type II (false negative) errors, as these are by far the most damaging to culture. Suppose the model wrongly classifies a non-toxic message as toxic. In that case, it is a minor annoyance to the moderator with the additional curation effort; in this case, less than 2 out of 100 flags, while wrongly classifying a toxic message as non-toxic, can be very detrimental to users and drive them away from the community. As the model is implemented, the model should be retrained regularly to reduce the false negative flags.

*Actor Contribution model*

The notion of a 'valuable' contribution policy is highly contextual (McGillicuddy et al. 2016). For this reason, the contribution policy was designed based on a series of consultations with moderators and senior participants. To ensure holistic coverage of the policy objective, a total of 21 attributes were included in the curation, retraining of the model to adapt continuously. 8 of the 21 attributes identified were deemed sufficiently objective to parametrize and model using the development framework:

| Classifier | precision | recall | f1-score | support |
|---|---|---|---|---|
| na | 0.89 | 0.93 | 0.91 | 156 |
| onboarding | 0.75 | 0.9 | 0.82 | 10 |
| knowledge_tcg | 0.57 | 0.5 | 0.53 | 16 |
| knowledge_fan | 0.67 | 0.6 | 0.63 | 10 |
| knowledge_crypto | 0.5 | 0.25 | 0.33 | 4 |
| content | 0.71 | 0.71 | 0.71 | 7 |
| moderation | 0 | 0 | 0 | 1 |
| suggestion | 0.5 | 0.29 | 0.36 | 7 |
| accuracy | | | 0.83 | 211 |
| macro avg | 0.57 | 0.52 | 0.54 | 211 |
| weighted avg | 0.82 | 0.83 | 0.82 | 211 |
| Table 4. Moderation model results | | | | |

The contribution model proved complex due to the contextual and subjective interpretation of community values. We initially tested a zero-shot prompt describing the desired contributions, which did not result in the ability classify contributions satisfactorily. We then extended the prompt to a broader taxonomy of what is considered contributions, fine tuning an ada model on the results. This also yielded suboptimal results. Finally, we extended the context window by including the previous messages in the conversation, to give the model a broader purview. While the results clearly do not match the intent and moderation models, we interpret the preliminary results as an indication that community moderation is feasible.

## Discussion

This short paper reports preliminary results towards the research question "to what extent can LLMs be used in moderating online communities?". We demonstrate that training student models, can yield results consistently performing in the upper quartile, indicating that LLM based student models will likely perform well in assisting, or even replacing, human moderators in online for and metaverse gaming experiences.

The models perform best when put to work on less context-sensitive tasks. This is especially evident when comparing the performance of the intent and moderation models to the contribution model. A shortcoming of the contribution model specifically, is that there is a limited amount of contribution data to train and validate on. As evident (Table 4.) 75% of the messages in the dataset were categorized 'na', meaning not a contribution. The remaining 25% was spread thinly across 8 categories. Increasing the amount of training and the volume of test data will likely improve these results significantly, provided that data reliability (Krippendorfs alpha) will improve over time as new sample data is generated.

GPT and prompt engineering presents a radical change in processing time for model development and performance. The ability to access pre-trained LLMs enabled us to produce results in weeks that would otherwise have taken months, at significant cost. The speed and ease at which these preliminary results were produced is indicative of what is to come, as the arms-race in artificial intelligence continues to proliferate.

While online community moderation is primarily an exercise in flagging and removing harmful content and classifying positive contributions, the practice still introduces ethical implications related to the use of artificial intelligence.

The methodology used in this short paper does not, in any meaningful way, account for the issues of perpetuating bias. In fact, one might argue that, while training student models limits the scope of the models' capabilities to their specific objectives, the model inherits any latent bias present in the teacher model. This issue can be addressed through the consistent re-training of student models with support from human moderators. These practices should envelop the student models into a supervisory system, in which moderators conduct proper oversight of the model's performance, until such a point that the model performs consistently within a set of boundaries.

## Conclusions and Future Work

This short paper presents preliminary results towards the task of moderating online communities. By training three task-specific student models aimed at identifying the intent of pseudonymous actors, identifying and moderating toxic behavior, and curating and promoting novel contributions, we find clear indications that these models will outperform existing solutions, if trained correctly. Future work on the project exhibited in this short paper entails re-training the models for datasets, collected from other Reality+ metaverse experiences to test the validity of the assumptions that the methodology presented here can be generalized. Additionally, we plan to extend the model with step-by-step distillation, a concept in which the LLM is asked to reason about its output. The rationale is then used as additional input in training the student model. This practice has shown promising early results in industry (Hsieh et al. 2023).

Further, from a cultural science perspective, the findings suggest that generative AI can be used to identify group formation around a common moral mission in borderless, pseudonymous, decentralized environments. Groups create knowledge and knowledge creates innovation, which is the foundation for economic systems (Hartley and Potts 2014). In this context, generative AI can become a new pathway for the automation of innovation to manage and expand social and cultural groups, thereby potentially bringing a new frontier for innovation in economic systems. Further work is required to define exactly how this innovation can be explored.

## Acknowledgements

This work is partially funded by a grant provided by the Danish Research and Innovation Council as administered and awarded by Copenhagen FinTech for the project Republic of Reality+. Additionally, the project received in-kind funding by Reality Plus ApS and Doerscircle Pte. Also, this research has received funding from the European Union's Horizon 2020 research and innovation programme, within the OpenInnoTrain project under the Marie Sklodowska-Curie grant agreement n°823971. Our work would not have been possible without the support from the Discord moderator team in Doctor Who Worlds Apart, the management team at Reality Plus and Doerscircle as well as the team at RMIT University Melbourne, Australia's Blockchain Innovation Hub. At the risk of leaving out others who deservedly should also be acknowledged by name, we are particularly indebted to Morten Rongaard and Professors Chris Berg and Jason Potts.

## References

Asatiani, A., Malo, P., Nagbøl, P. R., Penttinen, E., Rinta-Kahila, T., and Salovaara, A. 2021. "Sociotechnical Envelopment of Artificial Intelligence: An Approach to Organizational Deployment of Inscrutable Artificial Intelligence Systems," *Journal of the Association for Information Systems* (22:2), pp. 325–352.

Balasubramanian, N., Ye, Y., and Xu, M. 2022. "Substituting Human Decision-Making with Machine Learning: Implications for Organizational Learning," *Academy of Management Review* (47:3), Academy of Management Briarcliff Manor, NY, pp. 448–465.

Benbya, H., Davenport, T. H., and Pachidi, S. 2020. "Special Issue Editorial: Artificial Intelligence in Organizations: Current State and Future Opportunities," *MIS Quarterly Executive* (19:4), ix–xxi.

Brown, T. B., Mann, B., Ryder, N., Subbiah, M., Kaplan, J., Dhariwal, P., Neelakantan, A., Shyam, P., Sastry, G., Askell, A., Agarwal, S., Herbert-Voss, A., Krueger, G., Henighan, T., Child, R., Ramesh, A., Ziegler, D. M., Wu, J., Winter, C., Hesse, C., Chen, M., Sigler, E., Litwin, M., Gray, S., Chess, B., Clark, J., Berner, C., McCandlish, S., Radford, A., Sutskever, I., and Amodei, D. 2020. "Language Models Are Few-Shot Learners," *Advances in Neural Information Processing Systems* (2020-Decem).

Chen, J., Xu, H., and Whinston, A. B. 2011. "Moderated Online Communities and Quality of User-Generated Content," *Journal of Management Information Systems* (28:2), Taylor & Francis, pp. 237–268.

Elsbach, K. D., and Stigliani, I. 2019. "New Information Technology and Implicit Bias," *Academy of Management Perspectives* (33:2), pp. 185–206.

Glikson, E., and Woolley, A. W. 2020. "Human Trust in Artificial Intelligence: Review of Empirical Research," *Academy of Management Annals* (14:2), pp. 627–660.

Gou, J., Yu, B., Maybank, S. J., and Tao, D. 2021. "Knowledge Distillation: A Survey," *International Journal of Computer Vision* (129:6), pp. 1789–1819.

Hartley, J., and Potts, J. 2014. "Cultural Science : A Natural History of Stories, Demes, Knowledge and Innovation," *Cultural Science : A Natural History of Stories, Demes, Knowledge and Innovation*.

He, Q., Hong, Y., and Santanam, R. 2020. "Machine-Assisted Regulation, Online Participation and Human Moderation," *International Conference on Information Systems 2020*, pp. 0–9.

Hsieh, C.-Y., Li, C.-L., Yeh, C.-K., Nakhost, H., Fujii, Y., Ratner, A., Krishna, R., Lee, C.-Y., and Pfister, T. 2023. *Distilling Step-by-Step! Outperforming Larger Language Models with Less Training Data and Smaller Model Sizes*. (http://arxiv.org/abs/2305.02301).

Krippendorff, K. 2018. *Content Analysis: An Introduction to Its Methodology*.

Lindebaum, D., Vesa, M., and Den Hond, F. 2020. "Insights from 'the Machine Stops' to Better Understand Rational Assumptions in Algorithmic Decision Making and Its Implications for Organizations," *Academy of Management Review* (45:1), pp. 247–263.

McGillicuddy, A., Bernard, J. G., and Cranefield, J. 2016. "Controlling Bad Behavior in Online Communities: An Examination of Moderation Work," *2016 International Conference on Information Systems, ICIS 2016*, pp. 1–11.

Neelakantan, A., Xu, T., Puri, R., Radford, A., Han, J. M., Tworek, J., Yuan, Q., Tezak, N., Kim, J. W., Hallacy, C., Heidecke, J., Shyam, P., Power, B., Nekoul, T. E., Sastry, G., Krueger, G., Schnurr, D., Such, F. P., Hsu, K., Thompson, M., Khan, T., Sherbakov, T., Jang, J., Welinder, P., and Weng, L. 2022. *Text and Code Embeddings by Contrastive Pre-Training*. (http://arxiv.org/abs/2201.10005).

Nickerson, R. C., Varshney, U., and Muntermann, J. 2008. *A Method for Taxonomy Development and Its Application in Information Systems*, pp. 11–12.

Parasuraman, R., and Manzey, D. H. 2010. "Complacency and Bias in Human Use of Automation: An Attentional Integration," *Human Factors* (52:3), pp. 381–410.

Purdy, M., Zealley, J., and Maseli, O. 2019. "The Risks of Using AI to Interpret Human Emotions," *Harvard Business Review* (18:2019), pp. 11–19.

Regner, F., and Schweizer, A. 2019. "NFTs in Practice-Non-Fungible Tokens as Core Component of a Blockchain-

Based Event Ticketing Application Blockchain View Project Digitalization Cases View Project."

Salem, M., Lakatos, G., Amirabdollahian, F., and Dautenhahn, K. 2015. "Would You Trust a (Faulty) Robot?: Effects of Error, Task Type and Personality on Human-Robot Cooperation and Trust," *ACM/IEEE International Conference on Human-Robot Interaction* (2015-March), pp. 141–148.

Schanke, S., Burtch, G., and Ray, G. 2021. "Estimating the Impact of 'Humanizing' Customer Service Chatbots," *Information Systems Research* (32:3), pp. 736–751.

Seeger, A. M., Pfeiffer, J., and Heinzl, A. 2021. "Texting with Humanlike Conversational Agents: Designing for Anthropomorphism," *Journal of the Association for Information Systems* (22:4), pp. 931–967.

"Stanford CRFM." 2023. (https://crfm.stanford.edu/2023/03/13/alpaca.html, accessed April 20, 2023).

Tschang, F. T., and Almirall, E. 2021. "Artificial Intelligence as Augmenting Automation: Implications for Employment," *Academy of Management Perspectives* (35:4), pp. 642–659.

Zhou, Z., Zhang, D., Xiao, W., Dingwall, N., Ma, X., Arnold, A. O., and Xiang, B. 2022. "Learning Dialogue Representations from Consecutive Utterances," *NAACL 2022 - 2022 Conference of the North American Chapter of the Association for Computational Linguistics: Human Language Technologies, Proceedings of the Conference*, pp. 754–768.